# Quantum Photonic Estimation of Refractive Index in 2-methyl-4-nitroaniline ElectroOptic Crystal

Hassan Kaatuzian
*Professor of Electrical Engineering Department*

*Amirkabir University of Technology*

Tehran – IRAN
hsnkato@aut.ac.ir

*Abstract—* In this study we'll try to simulate and estimate theoretically the index of refraction of an assymetric organic ElectroOptic compound called (MNA) or 2-methyl-4-nitroaniline. It is transparent in a wide range between 500nm-2000nm wavelengths. We assume a Quantum Photonic Model which is based on Bohmian Mechanics for Electron-Photon interactions. The photons interact with π-electrons of MNA molecules, when they pass in the atomic layers. During this electron-photon interaction the photon is temporarily annihilated and all of it's energy will be delivered into electron in orbit as kinetic energy in every layer. After a short while the photon will be recreated. With reasonable and precise calculations of these delays we may estimate theoretically Ordinary and Extraordinary refracive indices of MNA compound. Quarter wave thickness for MNA crystal has also been estimated according to phase retardation simulations. Attained results from this Quantum Photonic model - Montecarlo time domain method, is estimated to be very close to experimental quantities with errors always less than 6 percents.

## I. INTRODUCTION

Organic crystal compounds like NPP, MNA, MAP [1,2,3,15] ,have high linear and nonlinear optical properties, in comparison with their inorganic counterparts such as LiNbo3, GaP, BaTiO3, etc. These high figure of merits are because of their electrons delocalization and large electron-photon interactions. So, they have the potential for fast switching, large capacity information processing and high density data transmission which are the needs for modern optical communication systems. Very thin layers of these organic materials can now be fabricated as optical integrated circuits [13]. In this paper, we'll explain the method of theoretical estimation of index of refraction, in an organic crystal compound called : MNA with several microns thickness. To analyze such a microscopic device, the classical methods can no more be valid and more rigorous approaches are becoming necessary. The most direct approach is to simulate microscopic π-electron-photon interactions. Then by averaging the results, one can obtain the average index of refraction in a specific material. We call this quantum approach: Quantum Photonics. [4,10,14].

In section (II), we'll describe in brief, the molecular and crystal structure of MNA aromatic compound. A more precise estimation of index of refraction will be described in section (III), Quantum-Photonics model based on annihilation and recreation of photons in each interaction with π-electrons will be explained at section (IV). Also the microscopic time delay (τ) or retardation of photons traveling inside the crystal will be estimated. These retardations cause macroscopic indices of refraction, both in ordinary and extraordinary directions. We'll also have a conclusion section.

## II. MOLECULAR AND CRYSTAL STRUCTURE of MNA

MNA or 2-methyl-4-nitroaniline has monoclinic crystal structure. MNA, is a member of Benzene group, where methyl $(CH_3)$, amino $(NH_2)$ and nitro $(NO_2)$ substituted for three hydrogen atoms. The lattice constants of MNA monoclinic crystal are a=7.9, b=11.17 and, c=11.6 angstroms. The angle between "a" and "b" is β=137Degrees. MNA crystals melt at 131 degrees centigrade with density about 1.477 g/cm3. MNA is transparent at the wavelengths between 0.48 to 2 microns. Molecular structure and crystal structure of this compound have been demonstrated in figures (1-a) and (1-b). MNA molecule is a dipole and it's crystal structure is asymmetric [1, 5].

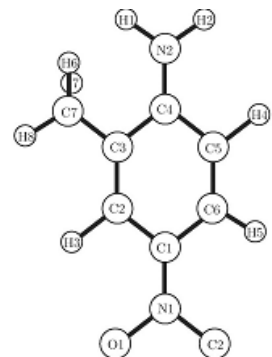

Fig. (1-a), MNA molecular structure.[4]

Some of the electrons associated with the carbonic double bonds are not actually localized between specific atomic pairs, but revolve around the entire ring. These electrons, known as π-electrons, are thus delocalized and their wave functions are of particular interest in this paper. To obtain the π-electron wave function, the Schrodinger equation may be solved. But this can not be done exactly and some approximated methods must be employed [6]. Since MNA is an organic compound from Benzene family, we can approximate the probability density function of π-electrons around six carbonic atoms.

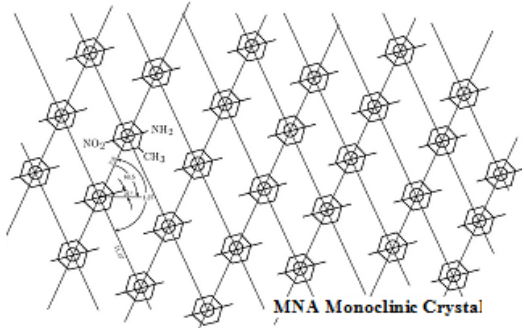

Fig. (1-b), Monoclinic crystal structure of MNA[4]

### III. MORE PRECISE ESTIMATION OF REFRACTIVE INDEX

Equation(1), shows the total time, which takes for photon to travel through a transparent media [4,10,12]:

$$T = \frac{L}{C_0} + \sum_{i=1}^{N} \tau_{di} \quad (1)$$

Where "L" is the material length. Travelling time of photon may be assumed as the sum of time that photon spends to pass through the intermolecular (inter-atomic) empty space ($\frac{L}{C_o}$) and photon-matter interactions ($\sum_{i=1}^{N} \tau_{di}$). The index of refraction is the ratio of $C_o$ (vacuum velocity of the light) over the average velocity of light in the medium for large values of N (number of interactions), so:

$$n = \frac{C_o}{C} = 1 + \frac{C_o}{L} \sum_{i=1}^{N} \tau_{di} \quad (2)$$

If now, we consider "$\tau_d$" as the mean retardation time per interaction and "d", as the mean free pass between two successive interactions, we have:

$$n = 1 + \frac{C_o}{d} \tau_d \quad (3),$$

(where: $\tau_d = \frac{\sum_{i=1}^{N} \tau_{di}}{N}$ ; and $d = \frac{L}{N}$)

During the interaction, as seen in fig. 2, at first, the photon, annihilates and gives it's energy to the electron in the lowest energy level and perturbs it. Since the energy of annihilated photon is not sufficient to transfer the electron to a higher allowed energy state, the perturbed electron returns to it's initial orbit after a transit time, which we call $(\tau_p)$, ultimately the photon recreated.

Therefore, the retardation time $(\tau_d)$ for a more precise estimation, may be considered as the sum of photon annihilation time $(\tau_a)$, electron perturbation time $(\tau_p)$ and photon recreation time $(\tau_r)$ (see fig.2) and [4,16]:

$$\tau_d = \tau_a + \tau_p + \tau_r \quad (4)$$

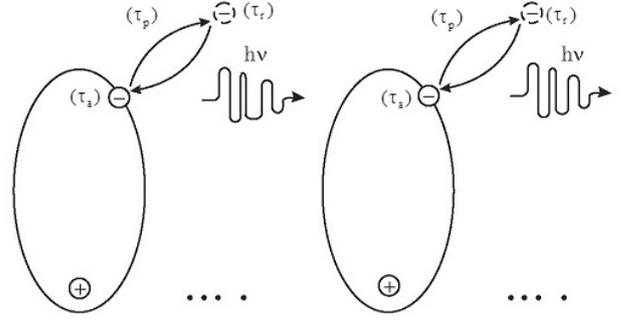

Fig. 2, Single photon-electron interactions with molecular layers of MNA based on Quantum Photonics and Bohm theory

Generally these quantities can be considered as a function of wavelength but, for simplicity as a first-order approximation, we consider $\tau_a$ and $\tau_r$ as negligible constants on the order of zepto-seconds. Since according to Quantum Photonics assumptions, $(\tau_p)$ will be in the range of atto-seconds [4].

### IV. QUANTUM PHOTONIC MODEL AND MONTECARLO METHOD

In this study, we've approximated, the position-probability of π-electrons of material "MNA" in space with an elliptic shape, centering the positive charges of molecule in one of the focal points of an ellipse with distinct eccentricity and π- electrons cloud in it's orbit. It's been shown that the probability of finding electron around the positive center in apogee position is more than it's probability in perigee. The probability density function (PDF) and cumulative distribution function (CDF) of π-electron elliptic orbital have already been calculated and estimated in [4]. To draw the curves in those figures, we use Kepler's law that says: "In an ellipse, The radius vector sweeps out equal areas in equal times" [4,6,7]. We assume the longer and shorter ellipse diameters "a" and "b" respectively. So that ellipse eccentricity ($\in$) can be obtained as [8]: $b^2 = a^2(1-\in^2)$. Now, if the total periodic time for sweeping the whole ellipse area ($\pi.a.b$) assumes to be "T", the time "t" that electron requires to sweep angle "$\theta$" around the ellipse orbit will be obtained[4,12]. In simulation, a random number generator will be predicted for electron angle ($\theta$)[11]. For a definite ellipse eccentricity, the instantaneous dipole length (r), will be calculated. If the photon's ordinary field vector is in angle($\theta$), regards to instantaneous dipole, the coulomb's force will be applied in angle ($\theta$) and the electron's velocity (V) will be: $V = \frac{V_o}{\cos(\theta)}$; $V_o$ is The radial component electron velocity. Now acceleration and velocity will be obtained:

$$a = K \frac{Q.q}{m_o.r^2} \quad (5)$$

$$V_o = \sqrt{\frac{2h\upsilon}{m}}.\cos(\theta) \quad (6)$$

Dividing "$V_o$" by "a", the radial acceleration, and the ordinary time retardation ($\tau_0$) will be estimated:

$$\tau_o = \frac{\sqrt{2h\upsilon m_o}.\cos(\theta)}{K.Q.q}.r^2 \quad (7)$$

So the Ordinary refractive index according to equation (2) will be simulated theoretically. A similar procedure may be applied to estimate extraordinary index of refraction.
But for photon's with extraordinary field vector, the angle between field vector and dipole will be: $(\pi/2 - \theta)$. So during a repeated procedure, similar to above mentioned, the extraordinary time retardation will also be computed. So That ordinary and extraordinary refractive indices will be estimated theoretically based on Quantum Photonics which itself is based on Bohmian Mechanics [14]. For example if we assume $(\upsilon = 5 \times 10^{14} Hz)$, Q=10.q, $q = 1.6 \times 10^{-19}$ coulomb, the time retardation ($\tau$) will be calculated on the order of several atto-seconds. This result according to Quantum Photonics has consistency with experiments [15]. Simulation results also show that there are both temporal and spatial fluctuations of refractive indices when ordinary or extraordinary polarized photons travel inside the crystal lattice. (see fig. 3) These fluctuations are around mean values, corresponds to macroscopic laboratory measurements for "MNA" crystal, with 50 microns thickness reported before. [6,15] Simulations have been repeated at different optical frequencies with wavelengths of $\lambda_1 = 6328$ Angstroms; $\lambda_2 = 5320$ Angstroms; $\lambda_3 = 10640$ Angstroms. Fig.4-a,b demonstrate the results of spatial phase retardation fluctuations in layer thicknesses below 100 Angstroms according to our simulations based on Quantum Photonics.

V. MOLECULAR LINEAR PERTURBANCES

If we apply an external transverse electric field to "MNA" crystal, in the range of several volts per micrometer, the physical shape of $\pi$-electrons cloud, which we have already approximated it with an elliptical orbit, will be deformed slightly. We would expect some noticeable variations in nanoscopic time delay parameters ($\tau_o \& \tau_e$)

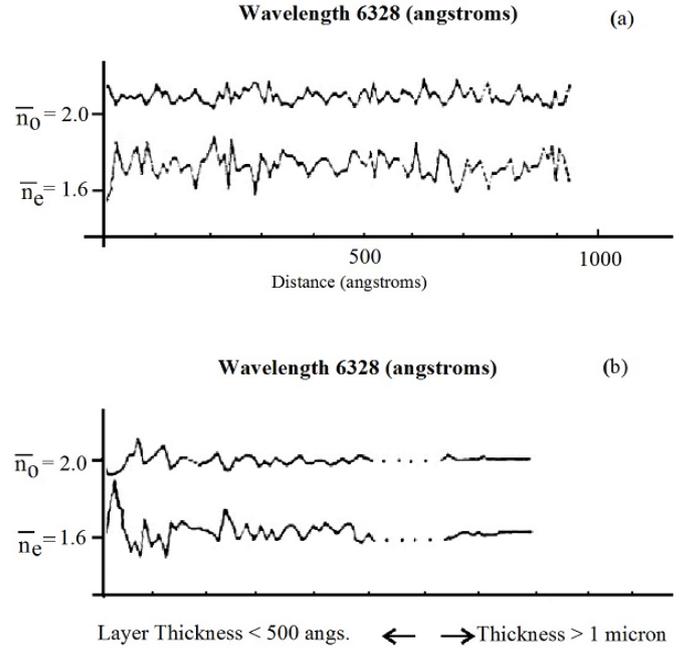

Fig. 3 (a)- Instantaneous fluctuations of ordinary (no) and extraordinary (ne) refractive indices simulated for MNA, at wavelength 6328 Angstroms. In (b), it's been shown that fluctuations of no and ne, in layer thicknesses below 500 Angstroms is considerable. But in thicknesses above micron, it's the same as experimental measurements. [15]

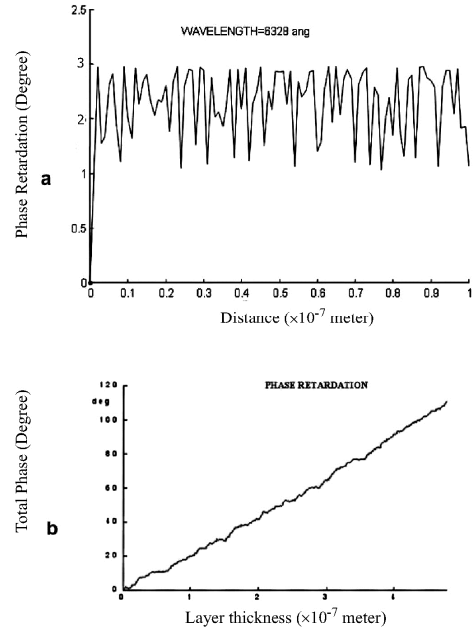

Fig. 4. (a) Spatial phase retardation fluctuations in every 100 Angstroms thickness of MNA monoclinic crystal simulated. (b) First quarter wave thick ness for MNA at wavelength: 6328 Angstroms and ellipse eccentricity of 0.8.

in case of applying external field. the simulation results have shown successfully the expected linear variations of phase retardation in terms of applied electric field. This is Quantum- Photonics interpretation for an important optical effect called in macroscopic scales: Electro Optics (E.O.) effect. (see fig. 5)

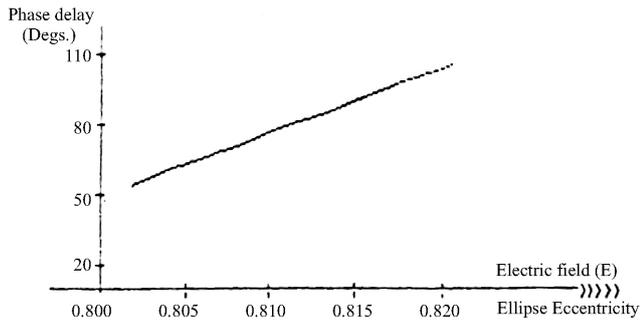

Fig.5, Montecarlo simulation of Linear Electro-Optic Pockles effect for MNA monoclinic crystal.

The results of this simulation can be successfully compared with experimental measurements reported before [15].

## VI. CONCLUSION

Diagrams of fig.3 and fig.4, demonstrate the results of simulation and estimation of ordinary and extraordinary refractive indices and also spatial phase retardation in terms of layer thickness. There exist fluctuations around mean values, corresponds to macroscopic laboratory measurements for "MNA" crystal, with 50 microns thickness reported before [15]. Simulations according to fig.3 have been repeated in other optical frequencies with wavelengths of $\lambda_1 = 10640$ angstroms and $\lambda_2 = 5320$ angstroms with very good similarities in comparison with experiments. As can be seen the maximum error is less than 6 percent at a large variations of photons energy (wavelength). These results show that analysis using Quantum-Photonics model not only delivers a visual perception about photon-electron interaction but also it delivers rather precise and valid results suitable for modern technological applications. The fluctuations of first quarter wave thickness, have also been simulated for theoretical estimation of phase retardation. That fluctuates are around 0.4 micrometer for MNA which is it's mean value in measurement reported [15]. See also fig. 4.